\documentclass[%
  aip,
  amsmath,amssymb,
  reprint,%
 ]{revtex4-1}

\usepackage{dcolumn,amsmath}
\usepackage{graphicx}
\usepackage{bm}
\usepackage{hyperref}

\usepackage[utf8]{inputenc}
\usepackage[T1]{fontenc}

\begin{document}

\title{The role of QED effects in transition energies of heavy-atom alkaline earth monofluoride molecules: a theoretical study of Ba$^+$, BaF, RaF and E120F}

\date{24.08.2021}

\begin{abstract}
Heavy-atom alkaline earth monofluoride molecules are considered as prospective systems to study spatial parity or spatial parity and time-reversal symmetry violating effects such as the nuclear anapole moment or the electron electric dipole moment. Comprehensive and highly accurate theoretical study of the electronic structure properties and transition energies in such systems can simplify the preparation and interpretation of the experiments. However, almost no attempts to calculate quantum electrodynamics (QED) effects contribution into characteristics of neutral heavy-atom molecules have been performed. Recently, we have formulated and implemented such an approach to calculate QED contributions to transition energies of molecules [L.V.~Skripnikov, J. Chem. Phys. \textbf{154}, 201101 (2021)]. In this paper, we perform a benchmark theoretical study of the transition energies in the Ba$^+$ cation and BaF molecule. The deviation of the calculated values from the experimental ones is of the order 10 cm$^{-1}$ and is more than an order of magnitude better than the ``chemical accuracy'', 350 cm$^{-1}$. The achievement of such an agreement has been provided in particular by the inclusion of the QED effects. The latter appeared to be not less important than the high-order correlation effects beyond the coupled cluster with single, double, and perturbative triple cluster amplitudes level. We compare the role of QED effects for transition energies with heavier molecules -- RaF and E120F, where E120 is the superheavy Z=120 homolog of Ra.
\end{abstract}

\author{Leonid V.\ Skripnikov}
\email{skripnikov\_lv@pnpi.nrcki.ru,\\ leonidos239@gmail.com}
\homepage{http://www.qchem.pnpi.spb.ru}
\affiliation{Petersburg Nuclear Physics Institute named by B.P. Konstantinov of National Research Centre ``Kurchatov Institute'', Gatchina, Leningrad District 188300, Russia}
\affiliation{Saint Petersburg State University, 7/9 Universitetskaya nab., St. Petersburg, 199034, Russia}
\author{Dmitry V.\ Chubukov}
\affiliation{Petersburg Nuclear Physics Institute named by B.P. Konstantinov of National Research Centre ``Kurchatov Institute'', Gatchina, Leningrad District 188300, Russia}
\affiliation{Saint Petersburg State University, 7/9 Universitetskaya nab., St. Petersburg, 199034, Russia}
\affiliation{Saint Petersburg State Electrotechnical University, street Professora Popova, St. Petersburg 197376, Russia} 
\author{Vera M.\ Shakhova}
\affiliation{Petersburg Nuclear Physics Institute named by B.P. Konstantinov of National Research Centre ``Kurchatov Institute'', Gatchina, Leningrad District 188300, Russia}
\affiliation{Saint Petersburg State University, 7/9 Universitetskaya nab., St. Petersburg, 199034, Russia}

\maketitle

\section{Introduction}

Alkaline earth monofluorides are an important class of diatomic molecules that can be used to test the nonconservation of symmetries in nature~\cite{Safronova:18,KL95}. At present, a few experiments on the barium monofluoride molecule, BaF, have been proposed. It is suggested to measure the electron electric dipole moment using BaF and thus test different models of new physics which imply violation of both spatial parity (P) and time-reversal (T) symmetries~\cite{BaF:2018,Kozlov:97,Abe:2018,Sunaga:2018,Hao:2019,Haase:2021}. Another suggestion is to experimentally study P-odd and T-even effects. One of the manifestations of such effects is the nuclear anapole moment~\cite{DeMille:2008,DeMille:2018,Nayak:2009}. In Ref.~\onlinecite{DeMille:2018}, a constraint on the fluorine anapole moment has been obtained using the $^{138}$Ba$^{19}$F molecule. The heavier analog of BaF is the radium monofluoride molecule, RaF. It is also considered as a prospective system to study T,P-odd, and P-odd effects~\cite{Kudashov:14,Isaev:2010,Isaev:2012,Ruiz:2019,Petrov:2020,Sudip:2016b,Borschevsky:13,Skripnikov:2021a,Skripnikov:2020e}. One of the features of these molecules is that they can be laser cooled to very low temperatures. Experiments with such molecules thus will benefit from the very large coherence time as the uncertainty of the measurement is inversely proportional to this time. In both molecules, the Franck-Condon matrix elements between the ground X~$^2\Sigma_{1/2}$ and the excited A~$^2\Pi_{1/2}$ states are highly diagonal~\cite{Kang:2016,Hao:2019,Ruiz:2019,Osika:2022}, allowing one to use this transition to laser-cool the molecule and increase the coherence time.  Recently, the first laser spectroscopy measurement on the radioactive molecule RaF has been reported~\cite{Ruiz:2019}. A further (super)heavier analog is the E120F molecule, where E120 is Z=120 element of the Periodic Table which has not been synthesized yet. Experiments with superheavy elements are much more complicated than those with their lighter analogs as one has to operate with extremely small concentrations, even with several atoms~\cite{Schwerdtfeger:2015}. Accurate predictions of transition energies in such molecules became even more important after the superheavy elements Factory had been launched~\cite{SHEFactory:2020}. Recently, it has been shown that three-atomic molecules that can be obtained by the substitution of the F atom with the OH and other groups have important advantages due to the $l$-doubling effect~\cite{kozyryev2017precision}. At the same time, in many cases electronic structure characteristics of these molecules such as the effective electric field acting on the electron electric dipole moment are close to the original alkaline earth monofluorides~\cite{Denis:2019}. It makes this class of molecules an important reference point for testing different theoretical approaches.

Previous theoretical studies of the BaF molecule and its analogs took into account both correlation and relativistic effects. However, we have not found any attempts to study the quantum electrodynamics (QED) effects. A few effective approaches called model Hamiltonians have been proposed to take into account QED effects in highly charged ions and single neutral atoms~\cite{Shabaev:13,Flambaum:2005,Pyykko:2003}. For example, recently, highly accurate predictions of the ionization potential, transition energies, and electron affinity have been made for these spherically symmetric systems~\cite{Schwerdtfeger:2017,Kaygorodov:2021,Guo:2021}. In some cases, the QED contribution is rather large, e.g. according to calculation~\cite{Kaygorodov:2021}, it reaches about 2.6\% of the total value of the electron affinity of Og~\cite{Kaygorodov:2021}. In Ref.~\onlinecite{Skripnikov:2021a} the expression for the model QED Hamiltonian has been proposed, which allows one to predict QED effects in molecules within the 4-component representation of the wave function. Another molecular implementation has been reported recently in Ref.~\onlinecite{Sunaga:2021}.

The BaF molecule has been thoroughly studied experimentally making it an ideal system to test new theoretical approaches. For this molecule, we perform benchmark calculations of the transition energies and compare them with the experimental values. We also perform such a study of the Ba$^+$ cation. Then we study and compare QED effects for heavier alkaline earth monofluorides: RaF and E120F.

\section{Theory}

As the first approximation for the molecular Hamiltonian we consider the Dirac-Coulomb (DC) approximation:
\begin{eqnarray}
 \nonumber
 H_{DC} &=& \Lambda_{+} \left[ \sum\limits_j\, ( c\, \boldsymbol{\alpha}_j \cdot {\bf{p}}_j + \beta_j c^2 
+V_{\rm nuc}(j) )
+ \sum\limits_{j<k}\,
\frac{1}{r_{jk}} \right] \Lambda_{+},
 \label{EQ:HAMILTONIAN}
\end{eqnarray}
where $\alpha$, $\beta$ are Dirac matrices, $V_{\rm nuc}$ is the potential due to the nuclear subsystem of the molecule, $\Lambda_{+}$ are projectors on the positive-energy states and summation is over all electrons. The leading two-electron correction to this Hamiltonian is the Gaunt term:
\begin{equation}
\label{HG}
 H_{G}
 =
 -
 \sum_{j<k}
 \Lambda_{+}
 \left(
  \frac{\left(
          \bm{\alpha}_j \cdot \bm{\alpha}_k 
      \right)}
      {r_{jk}}
 \right) \Lambda_{+}.
\end{equation}
This term is the dominant part of the Breit interelectron interaction for the systems under consideration~\cite{Eliav:1996,Skripnikov:2021a}. The QED effects which go beyond the Dirac-Coulomb-Breit approximation can be considered within the model Hamiltonian~\cite{Shabaev:13,Flambaum:2005,Pyykko:2003}. It includes the vacuum-polarization (VP) and self-energy (SE) terms:
\begin{equation}
\label{HQED}
 H_{QED}=H_{VP}+H_{SE}.
\end{equation}

The leading part of $H_{VP}$ can be represented by the Uehling potential. Several approximate expressions were derived to represent this potential. In the present paper, we use the formula from Ref.~\onlinecite{Ginges:2016} which takes a finite nuclear size effect into account and has a convenient form. The self-energy part of QED effects often gives a bigger contribution to the transition energy (see below) and is more complicated. In Refs.~\onlinecite{Shabaev:13,Flambaum:2005,Pyykko:2003,Thierfelder:2010,Schwerdtfeger:2017,LeimbachAt:2020,Dyall:2013} a number of expressions for the model SE operators has been suggested and analyzed for the atomic case. In principle, the model SE operator implies a scaling of the Lamb shift result for the Coulomb potential~\cite{Shabaev:13,Flambaum:2005,Pyykko:2003,Thierfelder:2010,Schwerdtfeger:2017,LeimbachAt:2020,Indelicato:1990,Tupitsyn:2013,Lowe:2013}. One of such expressions which can be easily applied for molecules has been suggested by one of us (L.S.) in Ref.~\onlinecite{Skripnikov:2021a}. It is close to the formulation~\cite{Shabaev:13} for the atomic case.  The possibility of scaling corresponding matrix elements in particular is possible due to the proportionality property of radial parts of heavy atoms wavefunctions having different principal quantum numbers $n$ and the same relativistic quantum number $\kappa=(-1)^{j+l+1/2}(j+1/2)$ \cite{Khriplovich:91,Titov:96,Titov:99,Shabaev:01a, Mosyagin:10a,Dzuba:2011,Titov:14a,Skripnikov:15b,Mosyagin:16,Skripnikov:16a,Skripnikov:2020e}. Let us consider functions $\widetilde{h}_{kljm}(\mathbf{r})$:
\begin{equation}
  \widetilde{h}_{kljm}(\mathbf{r})=\eta_{kljm}(\mathbf{r}) \theta(R_c-|\mathbf{r}|),
\label{hfuns}  
\end{equation}
where $\eta_{kljm}(\mathbf{r})$ are functions of the hydrogen-like ion and $\theta(R_c-|\mathbf{r}|)$ is the Heaviside step function, $R_c$ is a radius of a sphere with the origin at the heavy nucleus center. Due to a short-range property of the SE interaction $R_c$ is small and corresponds to radii of cutting functions in~\cite{Shabaev:13}. Matrix elements of SE over H-like functions $\int \eta_{nljm}^{\dagger}X\eta_{n'l'j'm'}d\mathbf{r}\ $ have been calculated and tabulated in Ref.~\onlinecite{Shabaev:13}. It is convenient to choose an orthogonal set of functions $h_{kljm}(\mathbf{r})$ which are linear combinations of $\widetilde{h}_{kljm}$ with a corresponding transformation of the $\int \eta_{nljm}^{\dagger}X\eta_{n'l'j'm'}d\mathbf{r}\ $ matrix elements. Now the model $X=H_{SE}$ operator can be written as follows (see Ref.~\onlinecite{Skripnikov:2021a} for details):
\begin{equation}
    H_{SE} \approx \sum_{kljm, k'l'j'm'} \frac{|h_{kljm}\rangle X_{kljm,k'l'j'm'} \langle h_{k'l'j'm'}|}{\langle h_{kljm}|h_{kljm}\rangle\langle h_{k'l'j'm'}|h_{k'l'j'm'}\rangle}.
\label{Xmod}    
\end{equation}

One of the problems of the construction of the operator in the form of (\ref{Xmod}) is a possible numerical linear dependence. In Ref.~\onlinecite{Skripnikov:2021a} the following scheme was proposed and implemented to overcome this problem. First one considers H-like functions $\eta_{nljm}$ and constructs corresponding functions $\widetilde{h}_{nljm}$ using Eq.~(\ref{hfuns}). Then one calculates the overlap matrix, having the scalar products of normalized functions $\widetilde{h}_{nljm}$ as matrix elements. As the final set of functions $h_{kljm}$ for Eq.~(\ref{Xmod}) one can choose eigenvectors of this matrix having eigenvalues larger than $10^{-6}$. Such a choice of functions reasonably solves the linear dependence problem. The procedure described above allowed us to use an analytic integration in molecular calculations. 

To compute the contribution of the QED effects we have added the operator~(\ref{Xmod}) to the molecular Hamiltonian at the correlation part of the electronic structure calculation. The difference between the total energies of a given electronic state of a given molecule obtained with and without the inclusion of this operator in the molecular Hamiltonian gives the QED contribution. 

\begin{table*}
\caption{Calculated transition energies for low-lying electronic states of the Ba$^+$ cation.
Deviations from the experimental values are given in the square brackets. All values are in cm$^{-1}$.}
\label{BaResults}
\begin{tabular*}{0.72\textwidth}{lrrrr}
\hline
\hline
                         & 5d $^2$D$_{3/2}$~~~~~~ & 5d $^2$D$_{5/2}$~~~~~~& 6p $^2$P$_{1/2}$~~~~~~ & 6p $^2$P$_{3/2}$~~~~~~ \\
\hline  
\multicolumn{5}{c}{This work:}    \\ 
55e-CCSD(T)              & 5115 & 5934 & 20302  & 22010  \\
Basis set correction     & -2   & -2   & 5      & 5     \\
CBS(L)                   & -104 & -100 & 3      & 3      \\
CCSDT-CCSD(T)            & 3    & 0    & -28    & -30    \\
CCSDT(Q)-CCSDT           & -6   & -4   & 10     & 12     \\
Gaunt                    & -68  & -93  & 26     & 7      \\
QED                      & -63  & -61  & -46    & -43    \\
Final (CCSDT(Q), DCG, QED)  & 4874 [0] & 5675 [0] & 20272 [-10]  & 21963 [-11]  \\
\\
\multicolumn{5}{c}{Other theory:}    \\ 
CP+SD, DCB, QED~\cite{Ginges:2015}   &  4812 [62]    & 5639 [36]     & 20153 [109]    & 21851 [101]     \\
FS-CCSD, DC ~\cite{Eliav:1996}       &  5268 [-394]  & 6093 [-418]   & 20396 [-134]   & 22103 [-151]    \\
MBPT-3, DC ~\cite{Guet:1991}         &  4688 [186]   & 5620 [55]     & 20995 [-733]   &  22742[-790]    \\
RCCSD, DC~\cite{Sahoo:2006}          &  5313 [-439]  &               & 20410 [-148]   & 22104 [-152]    \\
FS-CCSD, DC~\cite{Mani:2010}          &  4618 [256]   & 5434 [241]    & 20585 [-323]   & 22037 [-85]     \\
\\
Experiment~\cite{Nist2018}           & 4874          & 5675          & 20262            & 21952 \\
\hline
\hline
\end{tabular*}
\end{table*}

Dirac-Hartree-Fock and coupled cluster calculations up to the coupled cluster with single, double, and perturbative triple cluster amplitudes, CCSD(T), level were performed with the {\sc dirac}~\cite{DIRAC15} code. Higher-order coupled cluster calculations were performed with the~{\sc mrcc}~\cite{MRCC2020,Kallay:1} code. Fock-Space coupled cluster calculations have been performed within the 
{\sc exp-t} program package \cite{Oleynichenko_EXPT,EXPT_website}. Scalar relativistic correlation calculations to ensure the basis set completeness and to generate compact basis sets were performed using the {\sc cfour}~\cite{CFOUR} code. The code developed in Ref.~\onlinecite{Skripnikov:2021a} has been employed to calculate the QED contribution to molecular and atomic transition energies.

\section{Results and discussion}

Table \ref{BaResults} gives calculated values of the transition energies for several low-lying electronic states of the Ba$^+$ cation using several steps~\cite{Skripnikov:13a,Skripnikov:16b,Skripnikov:17b,Prosnyak:2020}. The main correlation contribution to the transition energies was calculated within the coupled cluster with single, double, and perturbative triple cluster amplitudes method, CCSD(T) using the Dirac-Coulomb Hamiltonian. All electrons were correlated and the energy cutoff for virtual orbitals included in the correlation calculation has been set to 10000 Hartree~\cite{Skripnikov:17a,Skripnikov:15a}. For these calculations, we have used the basis set which was obtained by extending Dyall's AEQZ basis set for Ba~\cite{Dyall:06}. This basis has been augmented by diffuse functions of $s-$, $p-$, $d-$ and $f-$ types; we have also reoptimized last 3 $g-$ type functions and added 4 $h-$ and 3 $i-$ type functions. To construct these $g-$, $h-$ and $i-$ type functions we have used the scheme of generating natural-like basis set within the procedure  described and developed in Refs.~\onlinecite{Skripnikov:2020e,Skripnikov:13a}. The final basis set LBas consists of 39 $s-$, 33 $p-$, 26 $d-$, 13 $f-$, 9 $g-$, 6 $h-$ and 3 $i-$ type uncontracted Gaussian functions, which can be designated as [39,33,26,13,9,6,3]. Next, we calculated basis set extension correction. For this, we have performed scalar-relativistic (SR) CCSD(T) calculations within the X2C approach~\cite{IliasX2C} and correlating 27 outer electrons. The extended basis set LBasExt [39,33,26,37,11,10,8] has been obtained from the LBas basis set by adding several $g-$, $h-$ and $i-$ type functions and replacement of $f-$ type function by an even-tempered set. As one can see from Table~\ref{BaResults} basis set extension correction is small. We have also calculated extrapolated contribution of higher harmonics (L>6) to treat a complete basis set limit, CBS, using the Fock-Space coupled cluster theory with single and double cluster amplitudes and Dirac-Coulomb Hamiltonian. Here we have extrapolated contributions of L=5,6,7 harmonics. To take into account correlation effects beyond the CCSD(T) level we have considered correlating contributions up to the coupled cluster with single, double, triple, and perturbative quadruple cluster amplitudes level, CCSDT(Q)~\cite{Kallay:6}. In these calculations, the SBas basis set has been used which is equivalent to the Dyall's AETZ basis set~\cite{Dyall:06}. At this stage, 19 outer electrons of Ba$^+$ have been included in the correlation treatment. The Gaunt interelectron interaction contribution has been calculated at the FS-CCSD level within the molecular-mean-field exact 2-component approach~\cite{Sikkema:2009}. All electrons were correlated within the LBas basis set. To calculate QED contribution, the model SE operator (\ref{Xmod}), as well as $H_{VP}$, have been added to the Dirac-Coulomb Hamiltonian before the correlation CCSD(T) stage using the LBas basis set and correlating all electrons of Ba$^+$. The theoretical uncertainty of the calculated transition energies
is expected to be lower than 20-25 cm$^{-1}$. It is mainly determined by the basis set imperfections as well as missed retardation part in the Breit interaction~\cite{Eliav:1996,Skripnikov:2021a}.

\begin{table*}[]
\caption{Calculated transition energies T$_{\rm e}$ of the low-lying electronic states of the BaF molecule in cm$^{-1}$. Deviation from the experimental value is given in the square brackets.}
\label{BaFResults}
\begin{tabular*}{0.9\textwidth}{lccccc}
\hline
\hline
                       & ~~~~~~~~A'~$^2\Delta_{3/2}$           & ~~~~~~~~A'~$^2\Delta_{5/2}$    & ~~~~~~~~A~$^2\Pi_{1/2}$  & ~~~~~~~~A~$^2\Pi_{3/2}$  & ~~~~~~~~B~$^2\Sigma^+_{1/2}$    \\
\hline                       
\multicolumn{6}{c}{This work:}    \\ 
37e-CCSD(T)            & 10963                      & 11378               & 11777            & 12420         &   14197      \\
Core(1s..3d)           & -3                         & -4                  & 0                & 3             &   2     \\
Basis set correction   & -66                        & -66                 & -28              & -28           &  -53  \\
CBS(L)                 & -75                        & -71                 & -21              & -22           &  -19  \\
CCSDT-CCSD(T)          & 32                         & 39                  & -48              & -64           &  -54   \\
CCSDT(Q)-CCSDT         & -18                        & -15                 & 18               & 26            &   35   \\
Gaunt                  & -53                        & -66                 & -6               & -17           &   -8   \\
QED                    & -37                        & -35                 & -30              & -29           &   -26  \\
Final (CCSDT(Q), DCG, QED)   & 10744 [-11]          & 11160 [-16]         & 11663 [-16]      & 12290 [-12]   &  14074 [-11]    \\
\\
\multicolumn{6}{c}{Other theory:}    \\ 
X2C-FSCC, DCG~\cite{Hao:2019}              &   10896   [-162]    &   11316   [-171]    &    11708 [-61]  &  12341 [-63]  & 14191 [-129]  \\
DFT(LDA), nonrelat.~\cite{Westin:1988}     &   7420    [3314]    &   7420    [3725]    &    9437  [2210] &  9437  [2841] & 12663 [1400]  \\
EPM, nonrelat.~\cite{Torring:1989}         &   11100   [-366]    &   11100   [45]      &    12330 [-683] &  12330 [-52]  & 14250 [-187]   \\
LFM,nonrelat.~\cite{Allouche:1993}         &   11310   [-576]    &   11310   [-165]    &    11678 [-31]  &  11678 [600]  & 13381 [682]   \\
CASSCF+MRCI, SR~\cite{Tohme:2015}          &   12984   [-2250]   &   12984   [-1839]   &    11601 [46]   &  11601 [677]  & 13794 [269]  \\
CASSCF+MRCI, SR+SOC~\cite{Kang:2016}       &   11582   [-848]    &   12189   [-1044]   &    12329 [-682] &  14507 [-2229] & 14022 [41] \\    
\\
Experiment~\cite{Barrow:1988}              & 10733.59            &     11144.59          & 11646.9         & 12278.2      &  14062.5   \\
\hline
\hline
\end{tabular*}
\end{table*}

For further application, we have calculated the correlation contribution of the $1s..3d$ electrons of Ba$^+$ within the FS-CCSD method using the Dirac-Coulomb Hamiltonian. The resulting contributions are -7~cm$^{-1}$, -9~cm$^{-1}$, 3~cm$^{-1}$ and 11~cm$^{-1}$ for 5d $^2$D$_{3/2}$, 5d $^2$D$_{5/2}$, 6p $^2$P$_{1/2}$ and 6p $^2$P$_{3/2}$ states, respectively. It has been also found that it is not important to include  $1s..3d$ core electrons in the correlation calculation to get the contribution of the Gaunt interaction. This can simplify more complicated molecular calculations where the symmetry of the system is lower than the atomic one.

Table \ref{BaFResults} gives calculated values of the transition energies for low-lying electronic states of the neutral BaF molecule.
The main correlation calculation of BaF has been performed within the CCSD(T) method using the Dirac-Coulomb Hamiltonian. $1s..3d$ electrons were excluded from the correlation treatment at this stage, i.e. 37 outer electrons have been correlated. In this calculation, we have used the LBasBaF basis set. It consists of [39,33,26,13,9,4,1] uncontracted Gaussians for Ba and corresponds to the LBas basis set with a reduced number of $h-$ and $i-$ type functions. For F the LBasBaF basis set is equivalent to the uncontracted Dyall's AETZ basis set~\cite{Dyall:2016} [14,8,3,1]. Correlation contribution of $1s..3d$ core electrons of Ba has been calculated within the FS-CCSD method and 
Dirac-Coulomb Hamiltonian. To take into account the extended basis set contribution we have used the LBasBaFExt basis set. It corresponds to the LBasExt basis set on Ba and the uncontracted Dyall's AAEQZ basis set [19,11,6,4,2] on F~\cite{Dyall:2016}. We have also calculated extrapolated contribution of higher harmonics (L>6) for Ba to treat a complete basis set limit, CBS, using the FS-CCSD method and the Dirac-Coulomb Hamiltonian similar to the atomic case. To treat high-order correlation effects up to the CCSDT(Q) level we have used the Dyall's AETZ basis set for both Ba and F~\cite{Dyall:2016,Dyall:06,Dyall:12} and included 27 outer electrons of BaF in correlation calculation. The Gaunt interelectron interaction contribution has been calculated at the FS-CCSD level of the treatment of correlation effects for 37 outer electrons within the molecular-mean-field exact 2-component approach~\cite{Sikkema:2009} and using the LBasBaF basis set.

Molecular transition energies given in Table \ref{BaFResults} were calculated as a difference between the total energies of the corresponding electronic states. For the ground electronic state X~$^2\Sigma_{1/2}$ we have used the value of the interatomic distance R$_{\rm e}=2.16$ \AA~while for the exited states the value R$_{\rm e}=2.18$ \AA~has been used. These values are close to the experimental values~\cite{Huber:79,Ryzlewicz:1980} for A~$^2\Pi_{1/2}$, A~$^2\Pi_{3/2}$ and B~$^2\Sigma^+_{1/2}$ states and to the theoretical equilibrium distances for A'~$^2\Delta_{3/2}$ and A'~$^2\Delta_{5/2}$ states obtained in the present paper. For these latter states, no experimental data are available. Table \ref{BaFSpectrResults} gives calculated and experimental spectroscopic constants for the considered electronic states of BaF. Calculations of the potential energy curves for X~$^2\Sigma_{1/2}$, A'~$^2\Delta_{3/2}$, A'~$^2\Delta_{5/2}$, A~$^2\Pi_{1/2}$ and A~$^2\Pi_{3/2}$ states have been performed within the scalar-relativistic X2C approach~\cite{IliasX2C} using the CCSD(T) method to treat electronic correlation effects for 37 outer electrons of BaF. For the B~$^2\Sigma^+_{1/2}$ state the CCSD-EOM-EA approach has been used. For these calculations, we have used the LBasBaF basis set where the uncontracted AETZ basis set for F has been replaced by the uncontracted AAEQZ one. Correction on the basis set superposition error has been taken into account. One can see good agreement with the experimental data where available.


\begin{table*}
\caption{Calculated spectroscopic constants for the ground and low-lying excited electronic states of BaF: equilibrium distance (R$_{\rm e}$), vibrational constant ($\omega_e$) and anharmonic vibrational frequency ($\omega_e x_e$).}
\label{BaFSpectrResults}
\begin{tabular*}{0.8\textwidth}{lllllll}
\hline
\hline
           & X~$^2\Sigma_{1/2}$           & A'~$^2\Delta_{3/2}$           & A'~$^2\Delta_{5/2}$    & A~$^2\Pi_{1/2}$  & A~$^2\Pi_{3/2}$ &  B~$^2\Sigma^+_{1/2}$    \\
\hline
\multicolumn{7}{c}{R$_{\rm e}$, \AA:}      \\
X2C-CCSD(T), SR, This work            & 2.167 & 2.196  & 2.196  & 2.183  & 2.183 & 2.215$^a$\\
\\
X2C-FSCC, DCG~\cite{Hao:2019}         & 2.177 &  2.207 &  2.205 &  2.196 &  2.195 &  2.222 \\
CASSCF+MRCI, SR~\cite{Tohme:2015}     & 2.204 &  2.229 &  2.229 &  2.197 &  2.197 &  2.234  \\
CASSCF+MRCI, SR+SOC~\cite{Kang:2016}  & 2.171 &  2.187 &  2.192 &  2.199 &  2.217 & 2.226  \\
Experiment~\cite{Huber:79,Ryzlewicz:1980} & 2.1592964(75) &              &              & 2.183        & 2.183   & 2.208     \\
           &               &              &              &              &              \\
\multicolumn{7}{c}{$\omega_e$, cm$^{-1}$:}      \\           
X2C-CCSD(T), SR, This work           & 465.4         & 436.1        & 436.1        & 438.4        & 438.4   & 420.2$^a$     \\
\\
X2C-FSCC, DCG~\cite{Hao:2019}         & 468.4 &  437.3 &  439.1 &  440.9 &  440.5 &  425.5  \\
CASSCF+MRCI, SR~\cite{Tohme:2015}     & 459.3 &  438.3 &  438.3 &  452.7 &  452.7 &  437.0  \\
CASSCF+MRCI, SR+SOC~\cite{Kang:2016}  & 474.1 &  446.3 &  423.3 &  456.7 &  417.7 &  421.7  \\
Experiment~\cite{Barrow:1988}         & 469.4 & 436.9  &  438.9 & 435.5  &  436.7 &  424.7        \\
                                &               &              &              &              &              \\
\multicolumn{7}{c}{$\omega_e x_e$, cm$^{-1}$:}      \\                             
X2C-CCSD(T), SR, This work            & 1.82         & 1.85        & 1.85        & 1.91        & 1.91    &  1.82$^a$    \\
\\
X2C-FSCC, DCG~\cite{Hao:2019}         & 1.83 &  1.84 &  1.82 &  1.90 &  1.87 & 1.81   \\
CASSCF+MRCI, SR+SOC~\cite{Kang:2016}  & 1.90 &  2.02 &  1.32 &  2.55 &  1.86 & 1.83  \\
Experiment~\cite{Bernard:1992}  & 1.83727(76)   & 1.833(27)    & 1.833(27)    & 1.854(12)    & 1.854(12) & 1.8524(37)   \\
                                &               &              &              &              &              \\
\hline
\hline
\end{tabular*}
\\
$^a$ Calculated within the X2C-CCSD-EOM-EA approach.
\end{table*}

Table \ref{BaBasis} gives the values of the transition energies between the ground and low-lying excited states of Ba$^+$ and QED contributions to these energies. These calculations have been performed within the Dirac-Coulomb Hamiltonian at the CCSD(T) level and correlating all electrons. Two basis sets have been used: the large basis set LBas and a smaller basis set SBas (see above). As one can see transition energies between the ground and 5d $^2$D$_{3/2}$ or 5d $^2$D$_{5/2}$ states are strongly dependent on the quality of the basis set used. In contrast, QED contribution in the proposed scheme~\cite{Skripnikov:2021a} has a much smaller dependence.

\begin{table}[]
\caption{Calculated values of excitation energies (EE) of Ba$^+$ within the Dirac-Coulomb Hamiltonian and QED contributions to these energies using the small (SBas) and large (LBas) basis set. The CCSD(T) method to treat electronic correlation effects has been used. All values are in cm$^{-1}$.}
\label{BaBasis}
\begin{tabular}{lrrrr}
\hline
\hline
         & 5d $^2$D$_{3/2}$ & ~~~5d $^2$D$_{5/2}$ & ~~~6p $^2$P$_{1/2}$ & ~~~6p $^2$P$_{3/2}$ \\
\hline         
EE/Sbas  & 6849  & 7624  & 20280 & 21978 \\
EE/Lbas  & 5115  & 5934  & 20302 & 22010 \\
QED/Sbas & -61   & -59   & -46   & -43   \\
QED/Lbas & -63   & -61   & -46   & -43   \\
\hline
\hline
\end{tabular}
\end{table}

\begin{figure}[]
\centering
\includegraphics[width = 3.4 in]{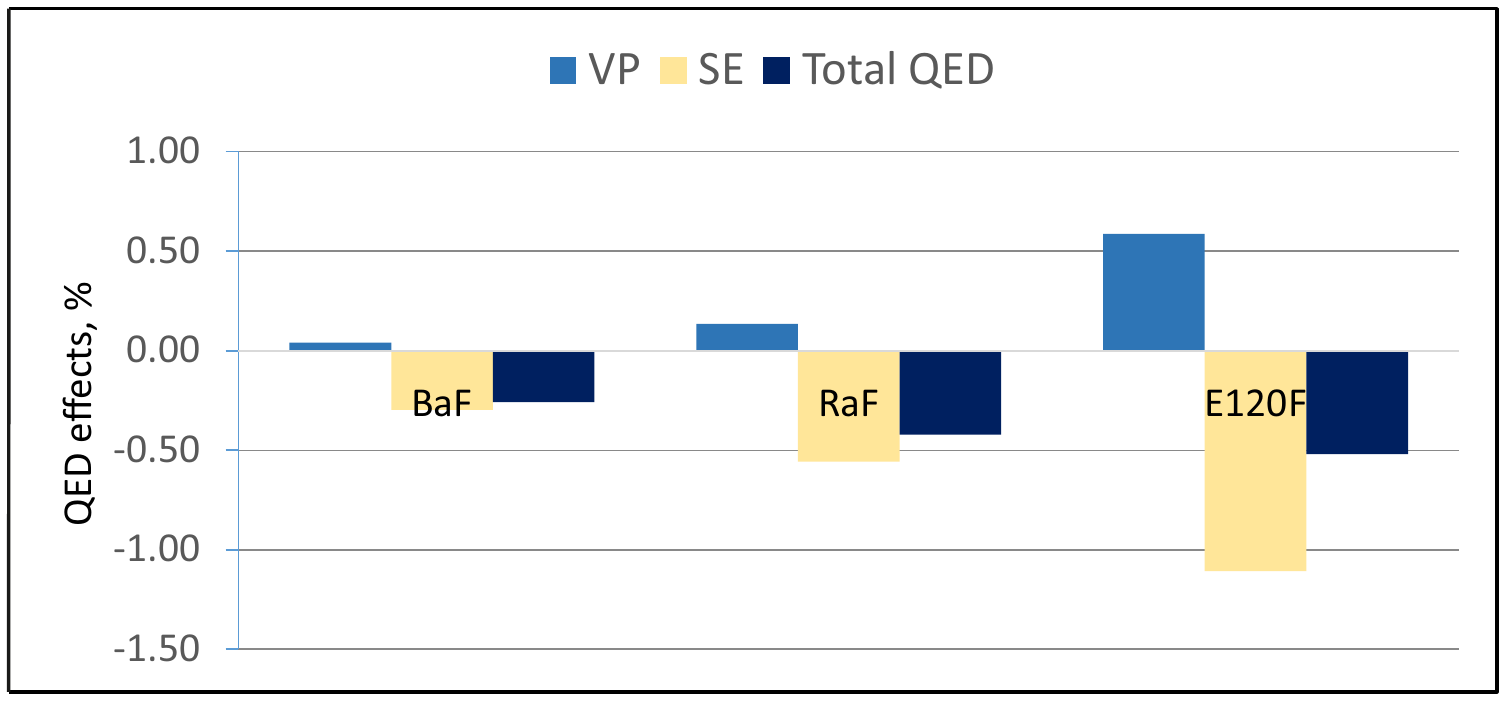}
 \caption{Contributions (in \%) of vacuum polarization, self-energy, and their sum to the transition energy between the ground $X~^2\Sigma_{1/2}$ and the excited A~$^2\Pi_{1/2}$ electronic states of BaF, RaF and E120F molecules.}
 \label{FigQED}
\end{figure}

As it was noted above, an important feature of the BaF and RaF molecules is the existence of transition between the ground $X~^2\Sigma_{1/2}$ and the excited A~$^2\Pi_{1/2}$ electronic states with highly diagonal Franck-Condon matrix elements~\cite{Kang:2016,Hao:2019,Ruiz:2019}. In the RaF molecule the A~$^2\Pi_{1/2}$ state is the first excited electronic state~\cite{Ruiz:2019}. In Fig.~\ref{FigQED} we compare the values of vacuum polarization and self-energy quantum electrodynamics effects contributions to the transition energy between the ground $X~^2\Sigma_{1/2}$ and the excited A~$^2\Pi_{1/2}$ electronic states for BaF and RaF molecules. To extend these data we have also studied QED effects in the superheavy analog of these two molecules -- the molecule of the monofluoride of the element with Z=120, E120F.

According to our study, the A~$^2\Pi_{1/2}$ electronic state is the first excited electronic state of E120F as in the RaF molecule and in contrast to the BaF molecule, where A'~$^2\Delta_{3/2}$ and A'~$^2\Delta_{5/2}$ states lie below the A~$^2\Pi_{1/2}$ state. It can be a manifestation of the relativistic stabilization of $s$ and $p$-electrons in Ra and E120. The determined equilibrium distance of the ground electronic state of E120F is 2.28~\AA. For the excited electronic state A~$^2\Pi_{1/2}$, R$_e$=2.27~\AA. These distances have been obtained from calculated potential energy curves taking into account correction on the basis set superposition error. These calculations have been performed within the two-component generalized relativistic effective core potential approach~\cite{Titov:99,Mosyagin:10a,Mosyagin:16,Mosyagin:17,Skripnikov:13a} at the CCSD(T) level and correlating 19 valence electrons of E120F. We have used the uncontracted natural-like~\cite{Skripnikov:13a} basis set [16,13,9,5,3,2] for E120 and the uncontracted AAEQZ basis set for F. The value of the transition energy between the ground $X~^2\Sigma_{1/2}$ and the first excited electronic state A~$^2\Pi_{1/2}$ of E120F is 17879~cm$^{-1}$, which is rather convenient for the experimental study as it lies in the optical range. This transition energy for E120F has been calculated at the FS-CCSD level within the molecular-mean-field exact 2-component approach taking into account the Gaunt interaction~\cite{Sikkema:2009} and using the uncontracted basis set [45,46,21,14,8,4,2] for E120 and uncontracted AETZ~\cite{Dyall:06} basis set for F.
This basis set for E120 has been obtained from the uncontracted Dyall's AAETZ basis set for Og~\onlinecite{Dyall:12}, where $s-$ and $p-$ type functions were replaced by the even-tempered set, while other type functions were partially replaced by the natural-like basis functions using the procedure from Ref.~\cite{Skripnikov:13a}.
There was no attempt to achieve a basis set limit as in the case of BaF and it will be studied elsewhere together with high-order correlation effects. On the other hand, the estimation of the QED effects in E120F should be accurate enough due to a rather modest basis set quality dependence of this contribution demonstrated above. For a comparison with other works, we have calculated the mean value of the SE operator (\ref{Xmod}) over the Dirac-Fock 8s orbital of neutral E120 atom having a sense of the self-energy contribution to the binding energy of this valence electron. Our value for this matrix element, -0.0333~eV, is in excellent agreement with the value, -0.0331~eV, obtained by Shabaev et. al. Ref.~\onlinecite{Shabaev:13}.

As one can see from Fig.~\ref{FigQED} and Tables ~\ref{BaResults} and ~\ref{BaFResults} the role of QED effects is not negligible even for BaF and Ba$^+$. One can also see that even a simplified treatment of the QED effects limited only by the inclusion of the vacuum polarization effects is not enough, as the self-energy contribution can be several times larger by absolute value and can have an opposite sign as demonstrated in Fig.~\ref{FigQED} for such element as Ba. Another observation is that the relative importance of VP and SE effects depends on the nuclear charge Z of the heavy element in the considered molecules: in the BaF molecule, the SE contribution is about 7 times bigger than the VP one for a given transition. For E120F the SE contribution is only two times bigger than VP one for this transition.

The Gaunt interaction contributes only 8 cm$^{-1}$ to the transition energy $X~^2\Sigma_{1/2} \to A~^2\Pi_{1/2}$ for the case of E120F. For BaF it is $-6$ cm$^{-1}$ (see Table \ref{BaFResults}) and for RaF this contribution is 5 cm$^{-1}$~\cite{Skripnikov:2021a}. Thus, for all three molecules, QED effect is several times larger than the Gaunt interaction contribution for the considered transition. This is not the case for other transitions, where QED and Gaunt effects give comparable contributions, see Table \ref{BaFResults}.

Finally, one can note, that QED effects contribution is bigger than the contribution of correlation effects beyond the CCSD(T) level for the considered electronic states of Ba$^+$, see Table~\ref{BaResults}. For the molecular case, high-order correlation effects contribution and QED contribution are comparable as can be seen from Table~\ref{BaFResults}. This is also true for Ra$^+$ and RaF molecule~\cite{Skripnikov:2021a}.

\section{Conclusions}

In the present paper, we have performed benchmark calculations of the excitation energies in the BaF molecule and Ba$^+$ cation. The uncertainty of these values is in many cases 1-2 orders of magnitude better than previous theoretical studies. At the present level of the electronic structure theory, QED effects have become important already for such elements as Ba (Z=56). Note, that new experiments to search for the manifestation of symmetry violating effects are suggested on radioactive molecules such as RaF and other systems for which experiments are complicated due to radioactivity and small amounts of required isotopes. Therefore, the role of accurate theoretical predictions becomes more and more important. We have shown a pilot application of the technique to study QED effects in heavy and superheavy molecules such as E120F and analyzed its features such as basis set dependence. It has been found that even for excitation energies of Ba$^+$ cation the contribution of QED effects can exceed the contribution of high-order correlation effects. The study of QED effects in molecules will be continued in further papers.

\begin{acknowledgments}
Electronic structure calculations have been carried out using computing resources of the federal collective usage center Complex for Simulation and Data Processing for Mega-science Facilities at NRC “Kurchatov Institute”, http://ckp.nrcki.ru/, and computers of Quantum Chemistry Lab at NRC ``Kurchatov Institute" - PNPI.

Study of the Gaunt and QED effects was funded by RFBR according to research project No. 20-32-70177. Dirac-Coulomb calculations were supported by the Russian Science Foundation (Grant No. 18-12-00227). SR calculations were supported by the foundation for the advancement of theoretical physics and mathematics ``BASIS'' grant according to Project No. 21-1-2-47-1. 
\end{acknowledgments}

\end{document}